\documentclass{sig-alternate}

\usepackage{balance}  
\usepackage{graphics} 
\usepackage{times}    
\usepackage{url}      

\makeatletter 
\def\url@leostyle{%
  \@ifundefined{selectfont}{\def\UrlFont{\sf}}{\def\UrlFont{\small\bf\ttfamily}}}
\makeatother
  
\usepackage{algpseudocode}
\usepackage{algorithm}
\usepackage{bm}
\usepackage{booktabs}
\usepackage{cite}
\usepackage{multirow}
\usepackage{subfigure}
\usepackage[usenames,dvipsnames]{xcolor}

\usepackage{tikz}
\usetikzlibrary{shapes,arrows,decorations.pathmorphing,backgrounds,fit,positioning,shapes.symbols,chains}

\definecolor{midblue}{rgb}{0.7,0.9,1}
\definecolor{orange}{rgb}{1,0.9,0.6}
\definecolor{lightgreen}{rgb}{0.6,1,0.6}
\definecolor{lightpink}{rgb}{1.0,0.7,0.9}
\definecolor{lightgray}{gray}{0.85}
\definecolor{lightred}{rgb}{0.8,0.5,0.5}

\newcommand{\comment}[2]{\begin{center}\colorbox{#1}{\parbox{0.85\linewidth}{\textit{{#2}}}}\end{center}}
\newcommand{\flavio}[1]{\comment{midblue}{Flavio: {#1}}}
\newcommand{\ju}[1]{\textcolor{blue}{#1}}

\DeclareMathOperator*{\argmin}{arg\,min}

\begin{document}



\setlength{\belowdisplayskip}{8pt} \setlength{\belowdisplayshortskip}{8pt}
\setlength{\abovedisplayskip}{8pt} \setlength{\abovedisplayshortskip}{8pt}

\permission{ECML/PKDD Discovery Challenge on Predictive Analytics}
\copyrightetc{Nancy, France, September 2014.\\Copyright is held by the auhtor/owner(s)}

\title{Improving the Effectiveness of Content Popularity Prediction Methods using Time Series Trends}

\newcommand{\superscript}[1]{\ensuremath{^{\textrm{#1}}}}
\def\sharedaffiliation{\end{tabular}\newline\begin{tabular}{c}}

\def\ao{\superscript{*}}
\def\at{\superscript{\dag}}
\def\ar{\superscript{\ddag}}

\def\dcc{Computer Science Department}
\def\ufmg{Universidade Federal de Minas Gerais}

\numberofauthors{1}
\author{
\alignauthor Flavio Figueiredo \quad Marcos Gon\c{c}alves \quad Jussara M. Almeida \\
\affaddr{Computer Science Department - Universidade Federal de Minas Gerais, Brazil} \\
\email{\{flaviov, mgoncalv, jussara\}@dcc.ufmg.br}
}

\maketitle

\begin{abstract}
We here present a simple and effective model to predict the popularity of web content. Our solution, which is the winner of two of the three tasks of the ECML/PKDD 2014 Predictive Analytics Challenge, aims at predicting user engagement metrics, such as number of visits and social network engagement, that a web page will achieve 48 hours after its upload, using only information available in the first hour after upload. Our model is based on two steps. We first use time series clustering techniques to extract common temporal trends of content popularity. Next, we use linear regression models, exploiting as predictors both content features (e.g., numbers of visits and mentions on online social networks) and metrics that capture the distance between the popularity time series to the trends extracted in the first step. We discuss why this model is effective and show its gains over state of the art alternatives. 


\end{abstract}


\section{Introduction}

With the ever-growing production of online content, characterizing and predicting user engagement (e.g., number of visits or social engagement such as Facebook likes) on web content may have multiple beneficial values such as: (1) understanding the human dynamics of information consumption; (2) supporting the decisions of content producers and providers on different tasks (e.g., marketing and content filtering); and, (3) understanding the physical processes that govern the growth of viewership on the Web. Several previous studies~\cite{Castillo2013,Figueiredo2011,Cha,Szabo2010} have characterized some of the factors that cause the popularity growth of different kinds of web content. Complementarily, various others~\cite{Ahmed2013,Pinto2013,Szabo2010,Nikolov2012} have focused on the the task of popularity prediction. We focus here on the latter task, aiming at predicting the popularity of a piece of content.

Popularity prediction is a difficult and important task since it mostly translates into income and profits for content providers, creators and consumers alike. For example, more visitors to a web page may lead to more ad-clicks and sales. Moreover, content provisioning to a large amount of users may require decisions such as geographical sharding of content to servers (due to the increased traffic). Thus, if planning is not performed correctly, longer latencies and loading times, and thus, fewer users may be expected. Finally, accurate and early predictions can lead to better services to the end consumer, such as search engine rankings~\cite{Radinsky2012}. 

We here present a simple, and yet effective, model for predicting the popularity of online content. More specifically, we present the winning model of two of the three tasks of the ECML/PKDD 2014 Predictive Analytics Challenge. In the challenge, different features related to the popularity of 30,000 web pages from 100 different hosts were provided. The goal of the challenge is to predict the popularity of 30,000 other pages from the same 100 hosts 48 hours after their upload. The features provided for the task were measured in the first hour after upload for each page.

Our model exploits the temporal features related to web pages (e.g., past visits and social engagement), as well as typical popularity (i.e., number of visits) time series trends which exist in the dataset. Such trends are extracted via unsupervised learning methods. Specifically, it combines the temporal features with features that capture the distances between the popularity time series for each web page and the extracted trends. We present a data characterization that motivates the design of our solution, and  show  the gains in prediction accuracy (ranging from 15\% to 27\%) when it is compared to  state of the art alternatives.

The rest of this paper is organized as follows. We formally describe the prediction problem and present the state of the art baseline methods in Section 2. In Section 3 we introduce our proposed solution, whereas our experiments and results are presented in Section 4. Finally, Section 5 concludes the paper.

\section{Background}

We start this section by defining the content popularity prediction problem (Section~\ref{subsec:ref}). In this definition, as throughout the rest of the paper, we refer to a particular piece of content as a web page\footnote{We here focus on web page popularity prediction, given the goal of the ECML/PKDD Challenge. However, our models are general and can be applied to other types of online content.}.  Next, we  discuss existing state of the art solutions used as baselines in our experimental study (Section~\ref{subsec:base}). 


\subsection{Problem Definition} \label{subsec:ref}

The  popularity prediction problem  we tackle can be stated as follows. Let $\mathcal{H}$  be a set of web hosts (e.g., \texttt{blogger.com}), where a single host $h \in \mathcal{H}$ is comprised of a set of pages,  $\mathcal{P}$ be the set of all pages,  where $p \in \mathcal{P}$ is a single page, and  $\mathcal{P}_{h}$ be the set of all pages from host $h$.
Moreover,  let $\mathcal{F}$ be a set of features associated with each page $p \in \mathcal{P}$, where each feature value is computed up to a certain \emph{reference time} $tr$ (e.g., $tr=$ 1 hour). Thus, using the set of features ($\mathcal{F}$) and the set of pages ($\mathcal{P}$), a matrix $\bm{X}_{tr}$ with $|\mathcal{P}|$ rows and $|\mathcal{F}|$ columns is defined for the values of features measured up to the reference time. Moreover, a row $\bm{x}_{p,tr}$ of the matrix $\bm{X}_{tr}$ defines the measurements for the given page\footnote{For simplicity, we shall identify rows using $p$.}. Using the measurements $\bm{X}_{tr}$, our goal is to predict the user engagement on each page up to a \emph{target time} $tt$ (where $tt > tr$). 


We here  focus on the following metrics of user engagement, referred to as   the response variables: number of visits $v_{p,tt}$, Facebook likes $f_{p,tt}$, and Twitter mentions  $m_{p,tt}$. All of them are cumulative measures, computed from page's upload up until time $tt$. We can then  define vectors of $|\mathcal{P}|$ rows for each response variable (e.g., $\bm{v}_{tt}$), or in more general terms, we can define a matrix $\bm{Y}_{tt}$ with three columns, one for each response variable:$ \bm{Y}_{tt} = [ \bm{v}_{tt}, \bm{f}_{tt}, \bm{m}_{tt} ]. $

With these definitions, the prediction task can be stated as a supervised machine learning task. Given a set of web pages for which both $\bm{X}_{tr}$ and $\bm{Y}_{tt}$ are available (the training set), our goal is to learn a function that maps $f(\bm{X}_{tr}) \rightarrow \bm{Y}_{tt}$. Ideally, such a function will generalize well for new pages not used in the training set. This function is usually defined as the {\it model}. The baseline methods, presented next, as well as our approach, introduced in Section 3,  explores linear regression method to learn the model.
Moreover, unless otherwise noted, we use a fixed $tr=1h$ as well as a $tt=48hs$ from now on, since these are the reference and target times defined in the Predictive Analytics Challenge. 

\subsection{Baseline Methods} \label{subsec:base}

One of the simplest prediction models, the Szabo-Huberman (SH) model \cite{Szabo2010}, defines one single feature for each page\footnote{The model was originally proposed for YouTube videos and Digg news.}, which is the number of visits measured up to the reference time $tr$. Using $tr$ = 1 hour, the SH model represents a single page as
$ \bm{x}_{p,1h} = <v_{p,1h}>$.
The SH model thus makes use of the following linear relation to provide predictions:

$$ log (1 + \bm{v}_{tt}) = log (1 + \bm{X}_{tr}) \theta. $$

Using linear regression, the parameter vector $\bm{\theta}$ (with only one cell in this case - $\theta$), is solved by minimizing:
$$ \min_{\bm{\theta}} || log(1 + \bm{X}_{tr}) \bm{\theta} - log(1 + \bm{v}_{tt}) ||^{2}_{2}, $$

\noindent where $|| \cdot ||^{2}_{2}$ is the squared $l2-norm$. The log transform is required given the linear correlations between $log(1 + \bm{v}_{tr})$ and $log(1 + \bm{v}_{tt})$ unveiled by the authors. The goal of this objective function minimizes the sum of squared errors on the log transformed data. We shall make use of the same objective since it is the one defined in the Predictive Analytics Challenge. However, we do note that in order to provide prediction in non-log transformed values, the authors suggest changing the linear regression objective by one based on the relative error, that is:
$$ \min_{\bm{\theta}} || (\bm{X}_{tr} \bm{\theta} - \bm{v}_{tt}) \circ \bm{v}^{-1}_{tt} ||^{2}_{2}. $$

\noindent where the inverse of a vector is defined as the cell-wise inverse, while $\circ$ is the cell-wise product (e.g., $\bm{x} \circ \bm{y} = <x_1y_1, ..., x_ny_n>$).

Pinto {\it et al.}~\cite{Pinto2013} extended the SH model by incorporating the whole history of the number of visits to the vector $\bm{x}_{p,tr}$. Using 5-minute time windows, the vector is defined as: 
$$\bm{x}_{p,tr} = <v_{p,5min}, v_{p,10min},v_{p,15min}, \cdots, v_{p,55min}, v_{p,1h}>.$$

Defining $\bm{v}_{p,tr}$ as the vector of visits measured in fixed length time windows (e.g., 5 minutes)\footnote{The model presented by Pinto {\it et al.}~\cite{Pinto2013} defines the amount of visits on each time window ($v_i$) not as cumulative (total views up to the window) as we do here, but actually as the amount of views gained in that window ($v_i - v_{i-1}$ in our notation). We found that using cumulative values lead to better results in terms of root mean squared error, thus we maintain our definition.}, the model above can be re-written as: $\bm{x}_{p,tr} = \bm{v}_{p,tr}.$
    
The same authors proposed a second model, called the RBF model,  which extends the set of features of each page  by adding {\it distance} features. Such distance features, measured using Radial Basis Functions\footnote{$RBF(\bm{x}, \bm{y}) = e^{||\frac{\bm{x}-\bm{y}}{\gamma}||^2_2}$, where $\gamma$ is a input parameter.}, are computed between the vector $\bm{v}_{p,tr}$ and a fixed number  $C$ of   vectors for other  pages, randomly selected from the training set. To avoid over-fitting, the authors suggest using ridge regression on the RBF model. Both the ML and RBF models were originally evaluated in terms of the relative errors, and not in terms of the log based regression as we do here.

Our last baseline is the model proposed  by Castillo {\it et al.}~\cite{Castillo2013}. In a very similar approach to the SH model, the authors also made use of a linear regression on log scales. However, instead of using one visit feature, the authors also explored social engagement features. Thus, a possible representation for a web page is: $$\bm{x}_{p,1h} = <v_{p,1h}, f_{p,1h}, m_{p,1h}>. $$

 In addition to these features, the authors also added other features, such as the entropy of tweets related to the web page. Since such features are unavailable in our dataset, we leave them out of the definition of the model. Finally, to mitigate issues of {\it multi-collinearity}, that is, correlation between predictors in the model, the authors suggest representing each page as:
$$\bm{x}_{p,1h} = <v_{p,1h}^2, f_{p,1h}^2, m_{p,1h}^2, v_{p,1h}f_{p,1h}, v_{p,1h}m_{p,1h}, f_{p,1h}m_{p,1h}>.$$

\noindent Since this model was initially proposed for news websites, we shall simply refer to it as the News model. 

\section{Our Approach} \label{subsec:meth}

Our approach combines the ideas described in the previous section with new features not explored by previous work. Moreover, as a novelty aspect, we make use of trend features extracted via  clustering of  visit time series. We first describe the features  we explore without considering these popularity trends (Section \ref{subsec:1}). Later, we discuss how we extract popularity trends and extend our model to include the distances between the popularity curve already observed of the page that is target of prediction and  the previously identified trends (Section \ref{subsec:2}).

\subsection{Mixed Model}
\label{subsec:1}

We borrow some of the ideas of the baselines by exploring the following temporal features for each page: (1) the time series of the number of visits to a page (each observation is recorded at each 5-minute time windows) - $\bm{v}_{p,tr}$; (2) two time series of user engagement which measure the number of Facebook likes - $\bm{f}_{p,tr}$, and the number of Twitter mentions -  $\bm{m}_{p,tr}$; (3) a time series of the average time each user spends on the page - $\bm{a}_{p,tr}$; (4) the weekday (e.g., Monday to Sunday)  and hour (e.g., 0 to 23) the page was created - $d_p$ and $c_t$. Moreover, we explore a single non-temporal feature which is the host to which each page belongs - $h_p$.

We encode the weekday and hour the page was created, as well as its host  in a binarized manner. That is, each value is represented by a sparse vector,  where  one cell, representing the given weekday (hour or host) has a value of one, and all other cells are zeroes. For example, a page uploaded on a Tuesday is represented as $<0, 1, 0, 0, 0, 0, 0>$. Thus, we   represent the weekday in which a page was uploaded as a vector $\bm{d}_p$, the hour as $\bm{c}_p$, and the host as $\bm{h}_p$. In this sense, each host, day of the week, and hour of the day become an \emph{indicator variable}.

With these features, one possible manner of representing each page is:
$$ \bm{x}_{p,1h} = <\bm{v}_{p,tr}, \bm{f}_{p,tr}, \bm{v}_{p,tr}, \bm{a}_{p,tr}, \bm{d}_{p,tr}, \bm{c}_{p,tr}, \bm{h}_{p,tr}>.$$

\noindent However, to mitigate multi-collinearity issues and to capture the behavior of hosts with non-linear popularity growth (discussed in the next section), we represent each page as:
\begin{align*}
    \bm{x}&_{p,1h} = <\bm{v}_{p,tr}, \bm{f}_{p,tr}, \bm{v}_{p,tr}, \bm{a}_{p,tr}, \\
                        &\bm{v}_{p,tr} \circ \bm{v}_{p,tr}, 
                        \bm{f}_{p,tr} \circ \bm{f}_{p,tr}, 
                        \bm{m}_{p,tr} \circ \bm{m}_{p,tr},
                        \bm{a}_{p,tr} \circ \bm{a}_{p,tr}, \\
                        &\bm{v}_{p,tr} \circ \bm{f}_{p,tr}, 
                        \bm{v}_{p,tr} \circ \bm{m}_{p,tr},  
                        \bm{v}_{p,tr} \circ \bm{a}_{p,tr},  
                        \bm{f}_{p,tr} \circ \bm{m}_{p,tr}, \\ 
                        &\bm{f}_{p,tr} \circ \bm{a}_{p,tr},
                        \bm{m}_{p,tr} \circ \bm{a}_{p,tr}, \\
                        &\bm{v}_{p,tr} \circ \bm{v}_{p,tr} \circ \bm{v}_{p,tr}, \\ 
                        &\bm{f}_{p,tr} \circ \bm{f}_{p,tr} \circ \bm{f}_{p,tr}, \\
                        &\bm{m}_{p,tr} \circ \bm{m}_{p,tr} \circ \bm{m}_{p,tr}, \\
                        &\bm{a}_{p,tr} \circ \bm{a}_{p,tr} \circ \bm{a}_{p,tr},
                        \bm{d}_{p,tr}, \bm{c}_{p,tr}, \bm{h}_{p,tr}>.
\end{align*}

We refer to this model as the Mixed model. The $\circ$ multiplications capture the same intuition as that of squaring the sum $(v_i + f_i + m_i)^2$ for each time window. Moreover, we also add the cubic terms (e.g., $v^3_i$) for the number of visits, Facebook likes, Twitter mentions and active time. To learn the model parameters we solve an linear regression task for each response variable.


\subsection{Mixed-Trend Model}
\label{subsec:2}

In order to capture the trend of each time series, we incorporate to the Mixed model features that capture the distance of the popularity curve of the target page measured during the reference time $tr$ to  given trends, which were previously identified using an unsupervised learning method. Specifically, we experiment with K-Means clustering~\cite{Hastie2009} and KSC clustering~\cite{Yang2011} to extract such trends from the training set. For each response variable, we define a matrix $\bm{T}_{tr}$, where each row is the time series of the response  for a given page:
%
%
$$ \bm{t}_{p,tr} = <\delta_{5min}, \delta_{10min}, \cdots, \delta_{55min}, \delta_{1h}>. $$

\noindent With the reference time of 1 hour, and a window length equal to 5 minutes, this matrix will have $|\mathcal{P}|$ rows and 12 columns. Each entry of the matrix, $\delta_i$, represents the {\em number visits} gained in that time window, i.e.,  $\delta_i = v_i - v_{i-1}$. 
We note that, using this matrix to extract trends is a common approach in the literature~\cite{Nikolov2012,Yang2011}.

The time series trends can be considered as the most common {\it shapes} of the different vectors $\bm{t}_{p,tr}$. Different techniques will extract shapes in different manners from a given training set. For example, the K-Means algorithm will group time series into $k$ clusters  according to the euclidean distance:
$$dist_{km}(\bm{t}, \bm{o})_{km} = || \bm{t} - \bm{o} ||^{2}_2. $$
 
 In contrast, the  KSC algorithm  groups times series based on a distance metric that is invariant of scale  in the popularity axis and shifts in the time axis \cite{Yang2011}. That is, two pages that have their popularities evolving according to similar processes (e.g., linear growth)  will be assigned to the same cluster by KSC, regardless of the popularity values. Also, two pages that have  stable popularity over time except for a  peak in a single window will also be clustered together, regardless of  the time when the peak occurred and the peak value. KSC is mostly a direct translation of the K-Means algorithm, except for the distance metric used, which is defined as: 
$$dist_{ksc}(\bm{t}, \bm{o}) = \displaystyle\min_{\alpha, q} \quad \frac{||\bm{t} - \alpha \bm{o}(q)||_2}{||\bm{o}||_2}.$$

\noindent where $\bm{o}(q)$ is the operation of shifting vector $\bm{o}$ by $q$ units. For a fixed $q$, the exact solution for $\alpha$, obtained by computing the minimum of $dist_{ksc}$, is: $\alpha = \frac{\bm{t}'\bm{o}(q)}{||\bm{t}'||_2}.$ The optimal value of  $q$ is found by considering all integers in the range of the size of the time series vectors (e.g., (-12,12)). 


It is important to note that, unlike KSC, K-Means is not scale invariant. Thus, in order to make the method invariant in terms of popularity we apply the following transforms. Initially, we apply a $log(1 + \bm{T}_{tr})$ to the time series matrices. Secondly, we z-normalize (zero mean normalization) each log transformed time series vector. While this approach will keep the popularity invariance, since time series will  have values in the same range, it does not tackle the time shifting invariant, as KSC does.  We also note that both K-Means and KSC receive $k$, the target number of clusters, as input.

Given a new page $p$ for which a prediction is to be made, we can compute the distances  between its popularity time series during the reference time  $tr$ and each  previously identified trend by simply computing the distances from $\bm{t}_{p,tr}$ to each cluster center (considering a fixed time window equal to $tr$), after clustering  in the training set is done using  either K-Means or KSC. Thus,
for each clustering method we can define a vector $\bm{s}_{p,tr}$ which includes the distances to the extracted trends. The Mixed-Trend model is thus the incorporation of these distance vectors into the Mixed model.




\begin{figure*}[ttt!]
 \centering
 \mbox{\subfigure[Visits]{\includegraphics{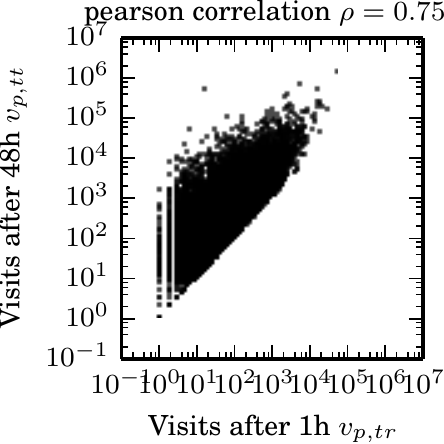}}}\hfill%
 \mbox{\subfigure[Facebook Likes]{\includegraphics{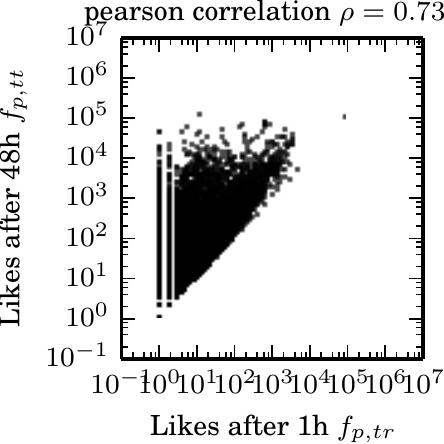}}}\hfill%
 \mbox{\subfigure[Twitter Mentions]{\includegraphics{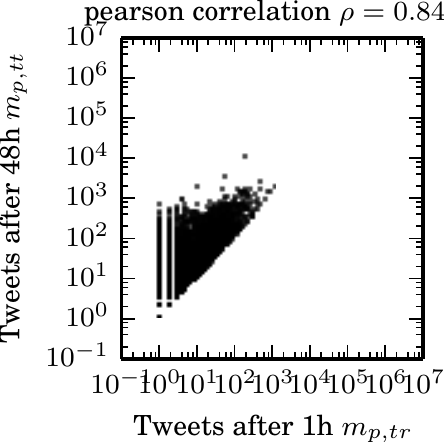}}}
 \caption{Correlations between the predictors number of visits $v_{tr}$, Facebook likes $f_{tr}$ and Twitter mentions (Tweets) $m_{tr}$ in one hour and their respective values after 48 hours. Each variable has been incremented by one due to log transformed x and y axes.}
 \label{fig:corr2h48}
\end{figure*}

\section{Experimental Evaluation}

We now discuss our experimental evaluation. We start by  discussing  how we trained  (i.e., parameterized) the models (Section \ref{subsec:cv}). Next, we provide some intuition on why our model works based on characteristics of the dataset (Section~\ref{sec:understanding}). We then compare both the Mixed and Mixed-Trend models and the baseline models (Section~\ref{sec:comparison}). 
The results discussed in this section are computed on the training set of the Predictive Analytics Challenge dataset, which consists of 30,000 web pages from 100 different hosts, each host with exactly 300 pages. We did not make use of the test set since the response variables $\mathbf{Y}_{tt}$ are not publicly available on the test set. Instead, we evaluate our models by employing Generalized Cross Validation, as described below.

\subsection{Model Parameterization} \label{subsec:cv}

For the SH, ML, News and Mixed models, model parameters  are learnt by the  regression method, i.e., by minimizing the  sum of squared errors on the log transformed data. However, for the RBF model, the parameter $\gamma$ (used by the RBF function), the regularization parameter of the ridge regression as well as the number $C$ of pages  selected to build Radial Basis Functions must be determined. Similarly, the number of clusters $k$ must be given as input for the Mixed-Trend model. 

Ideally, a temporal split of training and test sets would be performed to determine these parameters. However, given that the upload date of each page is not provided in the Predictive Analytics dataset, we decided to employ Generalized Cross Validation (GCV)~\cite{Hastie2009} to define the best parameter values. GCV is equivalent to leave-one-out cross validation (LOOCV). In LOOCV, one page per time is used to evaluate a model which is trained on the rest of  the pages. Thus, for each page, we compute the squared error between the predicted and real values. GCV computes the same squared error for each page without the need of manually splitting the dataset into train and test sets. Specifically, only one model is trained for the whole dataset, and the GCV measures compute the LOOCV error for every page~\footnote{The following website provides a good summary of GCV \url{http://robjhyndman.com/researchtips/crossvalidation/}}. When comparing different model parameters,  we measure the root mean squared error (RMSE) between the predicted and actual value for each page. The parameters with lowest RMSE are chosen.

For the Mixed-Trend model, we search for the best value of $k$  (i.e., number of clusters) in the  $[1, 100]$ range, finding it to be $k$=50 (for both K-Means an KSC algorithms) in all cases.  For the RBF model, we search for values of $\gamma$ and of the  ridge regularization parameter  considering the following options: $\{$0.001, 0.01, 0.1, 1, 10, 100, 1000$\}$. We also search for the best value of $C$ out of the options: $\{10, 50, 100\}$. The best parameter values were adopted in each case. When performing clustering, we make use of the entire dataset since we found that isolating a single page using the traditional LOOCV has little to no effect on our results.

We finally note that the SH, ML and RBF models are defined for a single engagement measure (e.g., number of visits). In order to evaluate these models for different engagement measures, we make the appropriate changes to the input features (e.g., changing from $v_{p,tr}$ to $f_{p,tr}$ or $m_{p,tr}$ in SH model).

\begin{figure*}[ttt!]
 \centering
 \mbox{\subfigure[Visits]{\includegraphics{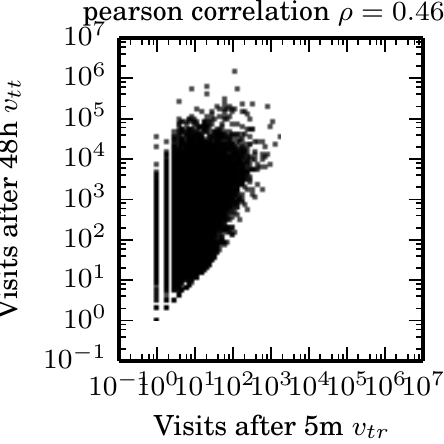}}}\hfill%
 \mbox{\subfigure[Facebook Likes]{\includegraphics{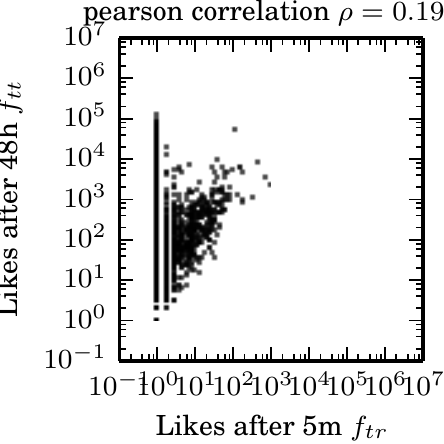}}}\hfill%
 \mbox{\subfigure[Twitter Mentions]{\includegraphics{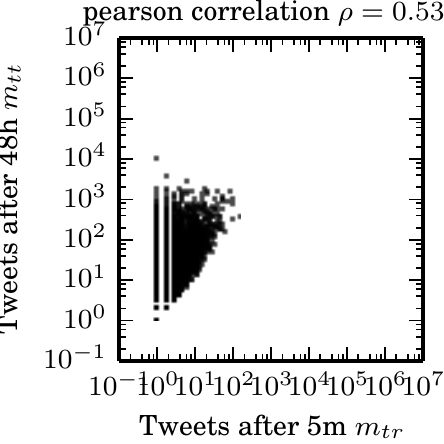}}}
 \caption{Correlations between the predictors number of visits $v_{p,tr}$, Facebook likes $f_{p,tr}$ and Twitter mentions (Tweets) $m_{p,tr}$ in 5 minutes and their respective values after 48h. Each variable has been incremented by one due to log transformed x and y axis.}
 \label{fig:corr5m48}
\end{figure*}

\begin{figure*}[ttt!]
 \centering
 \mbox{\subfigure[Visits vs Facebook Likes]{\includegraphics{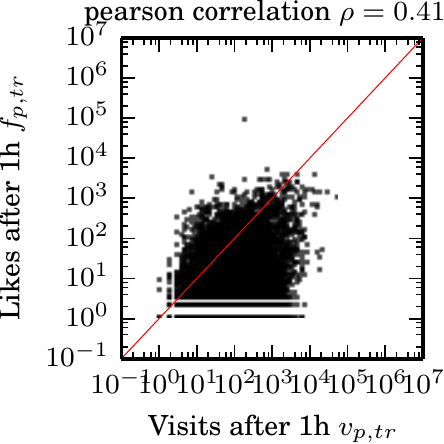}}}\hfill%
 \mbox{\subfigure[Visits vs Twitter Mentions]{\includegraphics{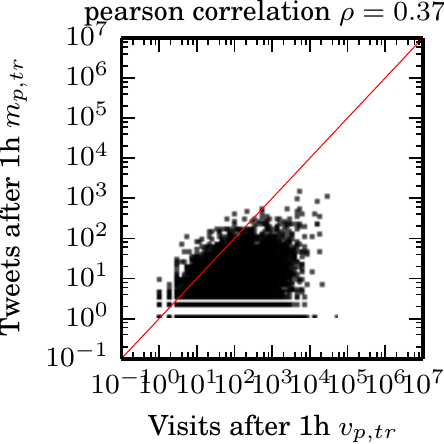}}}\hfill%
 \mbox{\subfigure[Facebook Likes vs Twitter Mentions]{\includegraphics{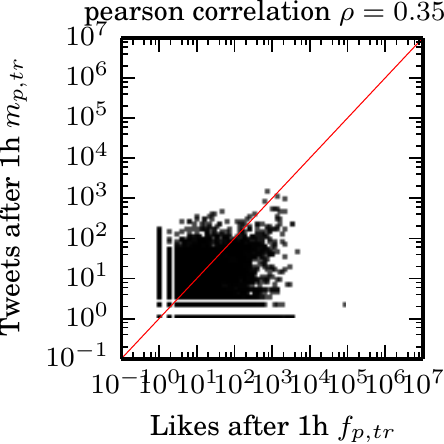}}}
 \caption{Correlations between the predictors number of visits $v_{p,tr}$, Facebook likes $f_{p,tr}$ and Twitter mentions (Tweets) $m_{p,tr}$. Each variable has been incremented by one due to log transformed x and y axis.}
 \label{fig:corr2hpairs}
\end{figure*}

\subsection{Data Characterization} \label{sec:understanding}

In order to motivate our model, we initially show the correlations between the user engagement metrics measured up to the reference time $tr$ and their respective values at the target time $tt$. Figure~\ref{fig:corr2h48} shows these correlations for the number of visits $v_{p,tr}$ (Figure~\ref{fig:corr2h48}-a), Facebook likes $f_{p,tr}$ (Figure~\ref{fig:corr2h48}-b) and Twitter mentions  $m_{p,tr}$ (Figure~\ref{fig:corr2h48}-c), using $tr=1$ hour. Note  that both axes of the graphs are in log-scales. Also, a value of 1 was added to each measure on each page (e.g., the axis for visits shows $log(1 + v_{p,tr})$).

The figure shows that a strong linear correlation in log scales (captured by the Pearson correlation coefficient $\rho$) exists for each engagement metric, as observed in \cite{Szabo2010}. Values of $\rho$ exceed $0.73$ for Facebook likes, reaching $0.84$ for Twitter mentions. Such strong positive correlations  motivate the use of linear regression methods to predict log-scaled engagement measures. However, the whole history of measures for each metric can also be useful to predict popularity values at $tt=48$ hours. This is exemplified in Figure~\ref{fig:corr5m48}, which shows scatter plots similar to those in Figure~\ref{fig:corr2h48}, but now assuming that $tr=5$ minutes. The figure shows that in some cases (such as visits and Twitter mentions), moderate  correlations (e.g. $\rho=0.46$ for visits and $\rho=0.53$ for Twitter mentions)  already exist even  very soon after the page was created. This motivates the use of the whole history of the measurements (e.g., $\bm{v}_{p,tr}$) and not only their final value at the reference time. 

We also looked at the correlations between engagement metrics.  Figure~\ref{fig:corr2hpairs} shows  that moderate correlations exist between every pair of metrics (e.g., $\rho$ of at least 0.35), which motivates our approach of multiplying  different metrics to mitigate multi-collinearity issues. More surprisingly, we find that there exists pages  that have more Facebook likes (and  Twitter mentions) than actual visits (points above the 45 degree line in each plot). This result indicates that not every like or tweet implies in a visit, and  suggests that measuring popularity on a single online social network service may be misleading, since people are not necessarily visiting the web pages. Finally, this result also suggests that we may not be able to completely rely on a single metric (e.g., Facebook likes) to predict the other (e.g., number of visits),  since only moderate correlations exist between them. 

Now that we have discussed the reasons behind using the whole history of the different metrics as predictors as well as behind our approach to deal with multi-collinearity,  we look into the motivation for also exploiting the host, day of the week and time of the day as predictors. Figure~\ref{fig:hosts} shows the correlations between number of visits at $tr$=$1$ hour and at $tt$=$48$ hours for  two hosts in our dataset. We note that host 68 (shown in black) has,  very similar values of $v_{p,tr}$ and $v_{p,tt}$ for most pages (i.e., most pages are on the 45 degree line). Such  finding implies that most pages of this host will not grow in views. In fact, if we train the SH model for this host only, it will find that the parameter $\theta$ has a value of $1.10$, that is, $log(1 + v_{p,tt}) = 1.10 log(1 + v_{p,tr})$. In contrast, host 3 shows a clear increase in popularity values for almost every page. In fact, the SH model, trained specifically for host 3, will capture the relationship between $v_{p,tr}$ and $v_{p,tt}$ as being $log(1 + v_{p,tt}) = 2.04 log(1 + v_{p,tr})$. This difference between hosts motivates the use of indicator variables (e.g., $\bm{h}_p$) to boost (positively or negatively) the general relationship  that exists in the whole dataset (see Figure~\ref{fig:corr2h48}) to relationships specific to the behavior of each host. Similarly, we can correct for the behavior for the different upload days ($\bm{d}_p$) and hours ($\bm{c}_p$). Finally, it further motivates the use of squared and cubic terms in the model to capture non-linear relationships between values at $tr$ to those at $tt$.

\begin{figure}[tt!]
 \centering
 \includegraphics{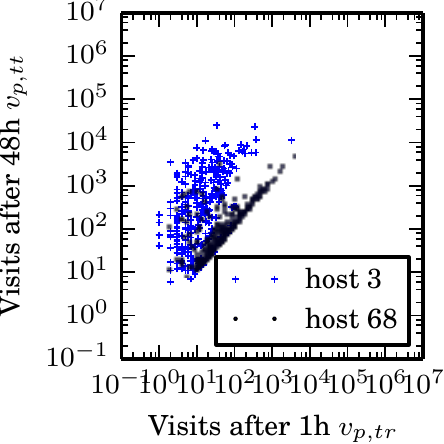}
 \caption{Correlation between $v_{p,tr}$ and $v_{p,tt}$ for selected hosts.}
 \label{fig:hosts}
\end{figure}

\begin{table*}[t]
\centering
\caption{Number of Features $|\mathcal{F}|$ and Prediction Results (Root mean squared error - RMSE)}
\begin{tabular}{lcccccccc|cc}
    \toprule
    & SH & ML & RBF & News & Mixed & Mixed-Trend & Mixed-Trend         &&& Mixed-Trend\\
    &    &    &     &      &       & KSC         & K-Means             &&& K-Means on $\bm{Y}_{tt}$\\
    \cline{2-11}
    $|\mathcal{F}|$ & 1 & 12 & 22 up to 112 & 60 & 347 & 397 & 397 &&& 397\\
    \cline{2-11}
    Visits           & 1.355 & 1.299 & 1.088 & 1.267 & 1.005 & {\bf 0.991} & {\bf 0.983} &&& 0.989\\
    Facebook Likes   & 1.835 & 1.793 & 1.534 & 1.525 & 1.390 & {\bf 1.383} & {\bf 1.380} &&& 1.378\\
    Twitter Mentions & 0.863 & 0.852 & 0.779 & 0.786 & 0.669 & {\bf 0.667} & {\bf 0.667} &&& 0.666\\
    \bottomrule
\end{tabular}
\label{tab:predresults}
\end{table*}

So far we have provided evidence that motivate our Mixed model. We motivate the Mixed-Trend model by showing in Figure~\ref{fig:examples} the evolution in the number of visits for two  web pages, selected from our dataset, that have similar popularity in terms of total number of visits. The figure shows that the numbers of visits of the two pages evolve over time according to very different processes. The web page shown in the black/solid line is steadily decreasing in popularity over time, whereas the web page in the blue/dashed line experiences a sharp increase in popularity 25 minutes after its upload. Such an example motivates the need for the Mixed-Trend model. Indeed, in \cite{Pinto2013} the authors argued that prediction accuracy could be improved by building specialized models for each popularity trend, although no attempt to learn  popularity trends and tackle such specialization was done. By incorporating the similarity of web pages to previously identified trends, as   proposed here,  we can effectively capture such differences in popularity curves, and thus improve prediction accuracy, as we shall discuss in the next section.

\subsection{Prediction Results} \label{sec:comparison}

We now discuss the prediction results in terms of the root mean squared error (RMSE) when measured using generalized cross validation (GCV). The results produced by all models, when using the best parameter values as discussed in Section \ref{subsec:cv},  are shown in  Table~\ref{tab:predresults}. On the table we also show the number of features of each model. Moreover, in last column of the table we also show the RMSE values obtained on the challenge server, that is, when measuring RMSE based on $\bm{Y}_{tt}$ and not using GCV.

Considering only the baselines, we find that the SH model performs worse than all other methods,  whereas the RBF model is the best baseline, except for predicting Facebook likes, for which the News model is the best baseline. More importantly, our proposed Mixed and Mixed-Trend models greatly outperform all baselines, for all three response variables. Moreover, by exploiting the distances to previously identified trends, the Mixed-Trend models, using either KSC or K-Means to extract the trends from the training set, also provides improvements over the simpler Mixed model, particularly for predicting number of visits.  Compared to the baselines, the improvements of the Mixed-Trend models vary from 15\% (for Twitter mentions against the RBF model) to 27\% (for the number of visits against the SH model). Finally, we note only marginal differences in RMSE (if any) between extracting trends using K-Means or KSC.  Thus, given the more scalable nature of K-Means~\cite{Yang2011}, we argue that the Mixed-Trend model using it as the trend extraction method is the most cost-effective solution.

Before concluding it is important to discuss whether over-fitting is occurring in our models. We argue that this is not the case based on three results. Initially, from the last column of Table~\ref{tab:predresults} we can see that the results for the Mixed-Trend K-Means model on the evaluation server test set is very close (and sometimes even smaller) than the one measured by GCV. Secondly, we also trained models using Ridge and Lasso regression~\cite{Hastie2009}, finding no improvements over the ordinary least squares linear regression we employ. Finally, we point out the result of Stone~\cite{Stone1977}, which shows that minimizing cross validated errors is asymptotically equivalent to minimizing Alkaike's Information Criterion (AIC). A similar result exists for linear models when using the Bayesian Information Criterion~\cite{Shao1993} (BIC). In order to avoid over-fitting both AIC and BIC penalize more complicated models. Thus, we also compared AIC and BIC values finding that the Mixed-Trend models always performs better than baseline approaches. These results indicate that on the Predictive Analytics Challenge dataset no over-fitting is occurring. However, it is impossible to generalize such a finding to any dataset. Thus, we point out that the use of regularized regression may be necessary on different datasets.

\begin{figure}[ttt!]
 \centering
 \includegraphics{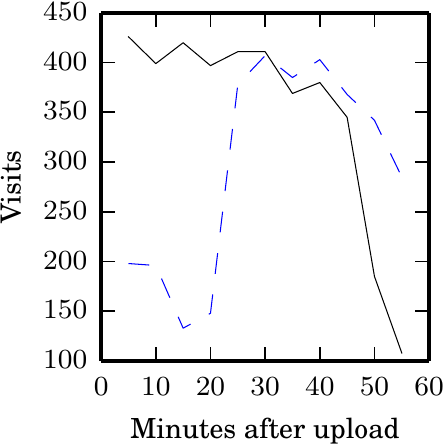}
 \caption{Popularity evolution of two selected pages.}
 \label{fig:examples}
\end{figure}

\section{Conclusions and Future Work }

We have presented a novel model that  exploits popularity time series (trends) and linear regression to  predict user engagement on web content. Three variations of the model were presented --- Mixed, Mixed-Trend KSC and Mixed-Trend K-Means --- together with a  data characterization that motivates their design. Our results show that our best model, the Mixed-Trend K-Means, provides gains in prediction accuracy ranging from 15\% to 27\% when compared with state of the art approaches\footnote{We note that all of our source code is available at: \url{http://github.com/flaviovdf/ecmlpkdd-analytics-challenge-2014}.}.

 Future work on popularity prediction includes addressing the tradeoff between accurate predictions and early predictions (i.e., how early a model can accurately predict the popularity of a piece of content) as well as  outlier detection (i.e., predicting that a content that has attracted little interest will suddenly burst in popularity).
 

\section*{Acknowledgment} 
This research is partially funded by a Google Brazil Focused Research Grant, the Brazilian National  Institute of Science and Technology for  Web Research (InWeb), CNPq, CAPES and Fapemig.

\bibliographystyle{abbrv}
{\small
\bibliography{bibs}
}

\end{document}